# Quantum Emitters in Hexagonal Boron Nitride: Principles, Engineering and Applications

*Thi Ngoc Anh Mai, Md Shakhawath Hossain, Nhat Minh Nguyen, Yongliang Chen, Chaohao Chen, Xiaoxue Xu, Quang Thang Trinh, Toan Dinh, Toan Trong Tran\**

Thi Ngoc Anh Mai, Md Shakhawath Hossain, Nhat Minh Nguyen, Toan Trong Tran
School of Electrical and Data Engineering, University of Technology Sydney, Ultimo, NSW, 2007, Australia
Email: trongtoan.tran@uts.edu.au

Yongliang Chen
Department of Physics, The University of Hong Kong, Pokfulam, Hong Kong 999077, China

Chaohao Chen
Department of Electronic Materials Engineering, Research School of Physics, The Australian National University, Canberra, Australian Capital Territory 2601, Australia

Chaohao Chen
ARC Centre of Excellence for Transformative Meta-Optical Systems (TMOS), Research School of Physics, The Australian National University, Canberra, Australian Capital Territory 2601, Australia

Xiaoxue Xu
School of Biomedical Engineering, University of Technology Sydney, Ultimo, NSW, 2007, Australia

Quang Thang Trinh
Queensland Micro, and Nanotechnology Centre, Griffith University, 170 Kessel Road, Nathan, QLD 4111, Australia

Toan Dinh

School of Engineering, University of Southern Queensland, Toowoomba, Queensland 4350, Australia

Toan Dinh
Centre for Future Materials, University of Southern Queensland, Toowoomba, Queensland 4350, Australia

**Abstract:**
Solid-state quantum emitters, molecular-sized complexes releasing a single photon at a time, have garnered much attention owing to their use as a key building block in various quantum technologies. Among these, quantum emitters in hexagonal boron nitride (hBN) have emerged as front runners with superior attributes compared to other competing platforms. These attributes are attainable thanks to the robust, two-dimensional lattice of the material formed by the extremely strong B-N bonds. This review discusses the fundamental properties of quantum emitters in hBN and highlights recent progress in the field. The focus is on the fabrication and engineering of these quantum emitters facilitated by state-of-the-art equipment. Strategies to integrate the quantum emitters with dielectric and plasmonic cavities to enhance their optical properties are summarized. The latest developments in new classes of spin-active defects, their predicted structural configurations, and the proposed suitable quantum applications are examined. Despite the current challenges, quantum emitters in hBN have steadily become a promising platform for applications in quantum information science.

## 1. Introduction

Quantum information science—a branch of study that seamlessly blends information theory and quantum mechanics—promises to revolutionize how we live, from ultra-secure transactions and parallel computing to the teleportation of information over distances. At the heart of such technologies are quantum hardware called quantum emitters that can release a single photon at a time, on demand. Traditionally, gas-phase platforms such as single atoms or single trapped ions were employed to generate single photons thanks to their superior optical properties. These platforms, however, suffer from drawbacks, including high cost, bulkiness, and complicated operations. Unlike these systems, most solid-state quantum emitters can operate at room temperature under ambient conditions and at a comparatively lower cost. Therefore, the past few decades have witnessed an explosion in research studies focused on exploring and utilizing new solid-state quantum emitters. To date, a wide variety of solid-state quantum emitters exist, from single molecules, colloidal and epitaxial quantum dots, to defect centers in diamond, silicon carbide, gallium nitride, carbon nanotubes, transition metal dichalcogenides (TMDs) and hexagonal boron nitride, as shown in **Figure 1a**.

Among these, quantum emitters in hBN have attracted significant research attention owing to their excellent optical properties and the unique hosting lattice. Hexagonal boron nitride has a honeycomb lattice structure similar to that of graphite. However, the hBN lattice is formed by the B-N bond that is both covalent and ionic, thanks to the relatively large differences in electron affinities of boron and nitrogen atoms. As a result, hBN is an indirect, wide bandgap material (~6 eV) that can host a variety of optically active defect centers. Such a wide bandgap value is critical for a host material since it can accommodate quantum emitters with various optical transition energies within its bandgap. In addition, hBN is a two-dimensional (2D) material that can be exfoliated layer-by-layer into single-atom-thick sheets of material, owing to the weak interplane van der Waals forces. This property allows for efficient extraction of the emission from the quantum emitters and excellent coupling to external photonic architectures, which will be discussed later in the text. Another advantage of hBN as a host stems from its extremely strong B-N bond that provides excellent chemical and physical protection for the embedded emitters. In the past decade, there have been several review articles about hBN quantum emitters. However, some discuss hBN quantum emitters as part of single-photon sources in 2D materials,[1-2] while others only give a brief summary of the quantum emitters.[3-4] Some other articles do give a detailed literature review on hBN quantum emitters.[5-6] These, nevertheless, need to be updated due to the fast-evolving nature of the topic. In this review, we curate the latest advancements in the fabrication and integration of hBN quantum emitters and provide in-depth perspectives and opinions based on our years of experience working with the material system. We also give an update on the rapid growth of the research topic based on our statistical bibliographic data.

The current review is arranged as follows. First, we introduce the fundamental properties of quantum emitters in hBN, including their excitation and detection schemes. We then discuss the current status of research on quantum emitters, followed by an overview of theoretical and experimental investigations on the origin of various defect families. Next, we delve into the fabrication of host materials and defect engineering of the quantum emitters using a range of strategies. We highlight some important works on integrating hBN quantum emitters into monolithic and hybrid photonic architectures. We subsequently overview the applications of quantum emitters in quantum information science, such as quantum sensing, quantum key distribution, and quantum communication. We discuss the robustness of the quantum emitters and suggest areas that can be improved. Finally, we conclude the review with an outlook on this active and exciting research topic.

## 2. Principles of quantum emitters in hBN

Similar to other solid-state emitters, quantum emitters in hBN behave like a two-level atomic system that is locked inside a lattice. Not only such implantation immobilizes the defect centers, it also promotes significant coupling between the defect centers and the lattice vibrations—the phonons. A typical emission spectrum of a quantum emitter in hBN is shown in **Figure 1b**, right panel. The spectrum entails a sharper peak at a shorter wavelength (higher energy), called a zero-phonon line (ZPL), and a broader peak at a longer wavelength (shorted energy), called a phonon-sideband (PSB). While the ZPL arises from the pure electronic transition and does not involve any phonon coupling, the PSB involves the coupling with the phonons of the defect and bulk lattice.

For this reason, the quantum emitters can be excited optically via three schemes. In the first scheme, the emitter can be pumped on resonance (**Figure 1b**, yellow dash box), where the excitation energy precisely matches that of the optical transition. Such an excitation protocol is the most efficient among the three methods and is similar to that for single atoms or single trapped ions in a vacuum. This is because of the huge absorption cross section at this excitation wavelength—the zero-phonon line. As efficient as it is, this excitation technique requires either advanced spectral filtering or orthogonal (cross) polarization. In the first method, only emitted photons falling into the PSB window are collected so that the majority of the back-reflected excitation laser can be rejected, rendering the high-purity single photons at the collection. The major drawback of this technique is that it only collects photons in the PSB, which are not very coherent. These photons are not useful in implementations requiring highly coherent photons, such as quantum communication or quantum entanglement. In the second method, the excitation and collected light are cross-polarized, for example, vertical and horizontal polarization for excitation laser and emitted photons, respectively. In this way, the residual laser can be substantially attenuated. In some cases, the attenuation can reach seven orders of magnitude or more, resulting in highly pure and coherent photons being harvested. This technique, however, does not come without a weakness. To enable effective filtering based on the different polarization of light, the sample flatness is critical to prevent scattering of the excitation laser, which in turn alters the initial polarization state of the excitation. When the sample is in powder/particle form below a few microns, diffusive scattering becomes significant, transforming linearly polarized light into elliptically polarized photons. This unwanted modification, in turn, results in a substantially lower signal-to-noise ratio. Therefore, the technique can only be applied for bulk crystals and not for samples with sizes smaller than a few microns.

In the second scheme (**Figure 1b**, green dash box), the emitter is excited with laser energies higher than the optical transition—also known as Stokes excitation. Such a protocol is allowed due to the efficient coupling with the local phonons. In this situation, the emitter emits a single or multiple phonons before generating a photon via the radiative decay channel—from the excited to the ground state. This scheme is the most popular choice for exciting defect centers in hBN since various laser wavelengths can be chosen, and the optical filtering procedure is straightforward due to the large spectral separation between the excitation laser and the emission signal. Unlike the resonant excitation procedure, Stokes excitation produces largely non-coherent photons, limiting the applicability of the emitters in various quantum technologies. In the third scheme (**Figure 1b**, red dash box), the emitter is pumped with laser energies lower than that of the optical transition—called Anti-Stokes excitation. Seemingly counterintuitive at first, the excitation is permitted thanks to the efficient phonon absorption process, which, in conjunction with the Anti-Stokes excitation, pumps the emitter into its excited state. Similar to the Stokes transition, Anti-Stokes excitation gives rise to incoherent photons. Interestingly, however, the optical transition follows the Arrhenius exponential scaling with temperature, making such an excitation valuable for thermal sensing applications. Once in the excited state, the electron decays to the ground state, emitting a photon. The average amount of time spent in the excited state is called the lifetime of the emitter. The shorter the lifetime is, the faster the emitter replenishes and emits another photon, hence the higher repetition rate. Quantum emitters in hBN typically possess lifetime values of ~1–3 ns, significantly shorter than well-known color centers in diamond such as the nitrogen-vacancy centers, whose have lifetime values of ~20 ns.[7] Not only do hBN emitters feature short excited-state lifetimes, but they also own impressive intrinsic quantum efficiencies (QEs)—quantities corresponding to the percentage of radiative relaxation. A recent study suggests a QE value exceeding 80% has been observed in hBN quantum emitters,[8] which is superior to most group-IV centers in diamond, such as the silicon-vacancy (SiV) or the germanium-vacancy (GeV),[9-10] where values of ~10–30% are typically observed. Such short lifetime, high quantum efficiency values, and the low refractive index of hBN (~1.8–1.9)[11] make the quantum emitters one of the brightest solid-state sources to date, with a repetition rate exceeding several MHz.[12-13]

The photon emitted from the hBN emitter can possess the wavelength of the ZPL or the PSB, depending on whether a phonon is simultaneously generated. For a typical hBN emitter, around 70% of the emission goes into the ZPL,[14-15] while the rest represents the PSB. This optical characteristic is another compelling feature of defect centers in hBN compared to competing platforms. To characterize the statistical nature of the emitted photons, the Hanbury Brown and Twiss (HBT) interferometer is typically employed (**Figure 1c**, left panel). The setup features a 50:50 (transmission:reflection) free-space or fiber-based beamsplitter that divides the collected light into two beams. Each beam is counted by a single-photon avalanche photodiode (SPAPD) capable of detecting extremely low light intensities. The two SPAPDs are time-tagged by an ultrafast time correlator with a typical resolution of a few picoseconds or better. The time-tagging scheme can be simplified as follows. The top SPAPD acts as a start switch upon the first photon arrival, which initiates the internal clock of the time correlator (**Figure 1c**, bottom panel). The bottom SPAPD then serves as a stop switch when the second photon is counted. The total delay time between the two consecutive photons is saved and binned into a histogram. After many cycles of collecting such time-correlating data, a histogram called the second-order correlation function is shown in **Figure 1c**, right panel. In the case of single photons, an antibunching dip value below 0.5, i.e. $g^{(2)}(0) < 0.5$ can be seen at the zero delay time—suggesting a sub-Poissonian statistic. Qualitatively, this indicates the low or zero probability of simultaneous detection of two photons at the two detectors. Theoretically, an antibunching value of zero is expected for a true single photon emitter. However, in real-world experiments, there are a few factors involved, such as background fluorescence, a second emitter located close to the excitation spot, or timing jitter due to the time-tagging errors. Realizing high-purity single photon emission is critical for various quantum applications that require the quantum interference effect. For quantum emitters in hBN, antibunching values below <0.1 were achievable in previous studies,[16-18] by using either thermal annealing treatment or laser writing to suppress contamination-induced background fluorescence. Higher purity single-photon emission ($g^{(2)}(0) = 6.4 \times 10^{-3}$) has also been demonstrated using a combination of confocal microcavity and ultrashort pulsed excitation to efficiently couple the emission into the cavity mode and eliminate background fluorescence and multiple photon generation.[19]

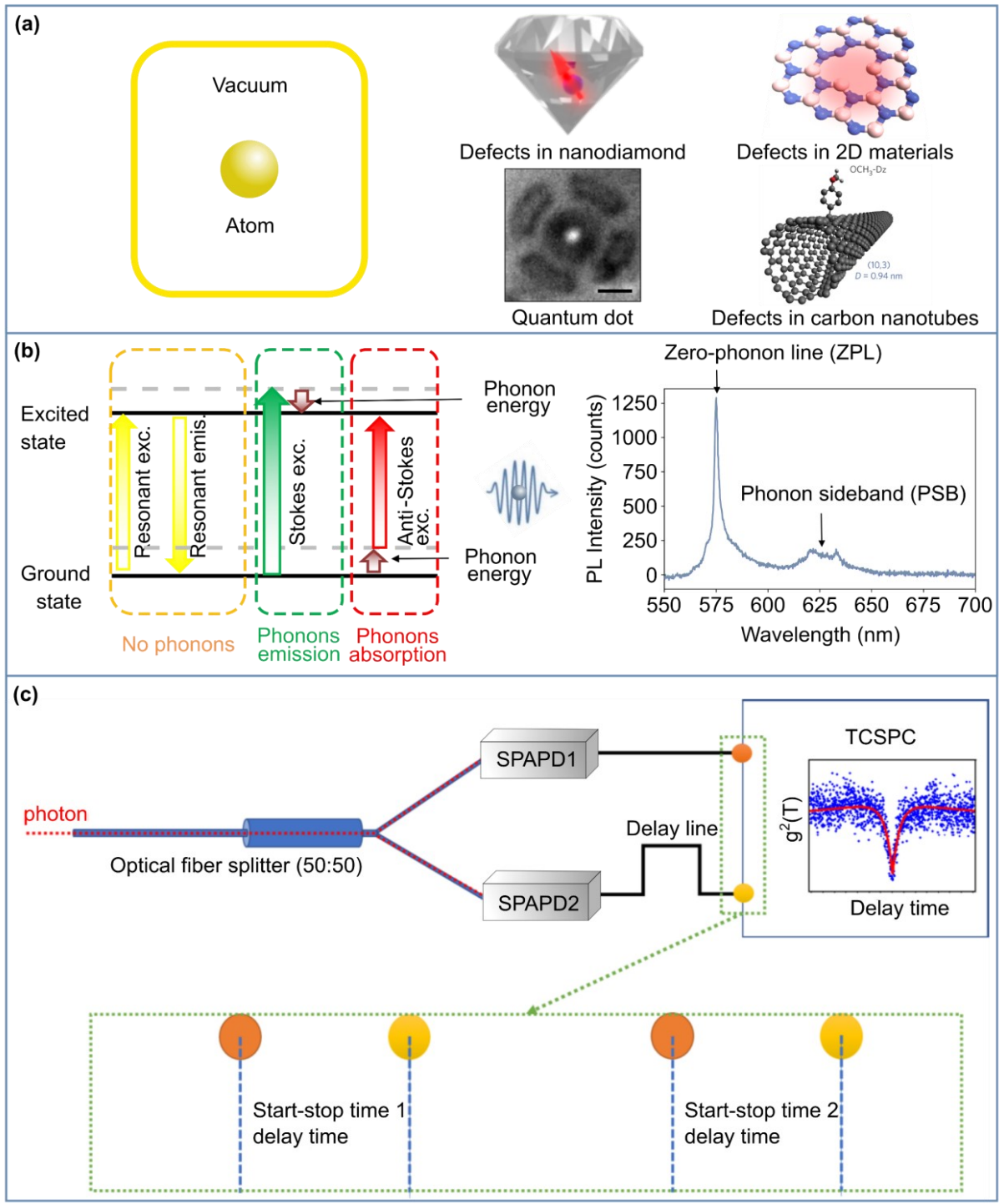

**Figure 1: Basic principles of quantum emitters in hBN. (a)** Classifications of quantum emitters: atom in vacuum (Left), and emitters in solid-state materials (Right): defects in nanodiamond, 2D materials (hBN and TMDs), carbon nanotubes and quantum dot **(b)** Excitation of single defect center in hBN and a representative spectrum. **(c)** The Hanbury Brown and Twiss setup for single-photon source measurement. SPAPD - single-photon avalanche photon detector, TCSPC - Time correlated single photon counting. Figure proposed defect structure in hBN adapted with permission.[12] Copyright 2016, Springer Nature. Figure

photoluminescence from a single quantum dot within the cavity (scale bar represents $5\mu m$) adapted under terms of the CC-BY 4.0 license.[20] Copyright 2015, The Authors, published by Springer Nature, Macmillan Publishers Limited. All rights reserved. Figure defect in carbon nanotubes adapted with permission.[21] Copyright 2017, Springer Nature.

## 3. Status of research on hBN quantum emitters

Research on hBN quantum emitters was pioneered by Tran and co-workers at the University of Technology Sydney, Australia, where they discovered the first quantum emitters operated at room temperature in 2016.[12] Since their first paper, the topic has gained significant traction in the research community, resulting in a sharp increase in publications and citations, exceeding 150 publications and 7000 citations per year, as shown in **Figure 2a-b**. To date, research groups worldwide have been increasingly active on the topic, especially in the United States, Australia, China, Germany, Japan, the United Kingdom, and other European countries (**Figure 2c**). Initially spearheaded by Australian researchers, the field has recently witnessed the largest growth in the number of publications in the US and China. With the recent discovery of new spin-active defect centers in hBN,[22-24] the research topic is expected to grow significantly in the following decades. A significant increase in research interest will be in quantum sensing thanks to the unique two-dimensional nature of the hexagonal boron nitride lattice that facilitates more efficient near-surface defect engineering than diamond or silicon carbide. Quantum communication, particularly quantum key distribution, will be another area of strong growth thanks to recent advances in optical engineering, allowing for integrating the quantum light sources and supporting electronics into an ultra-compact CubeSat design.

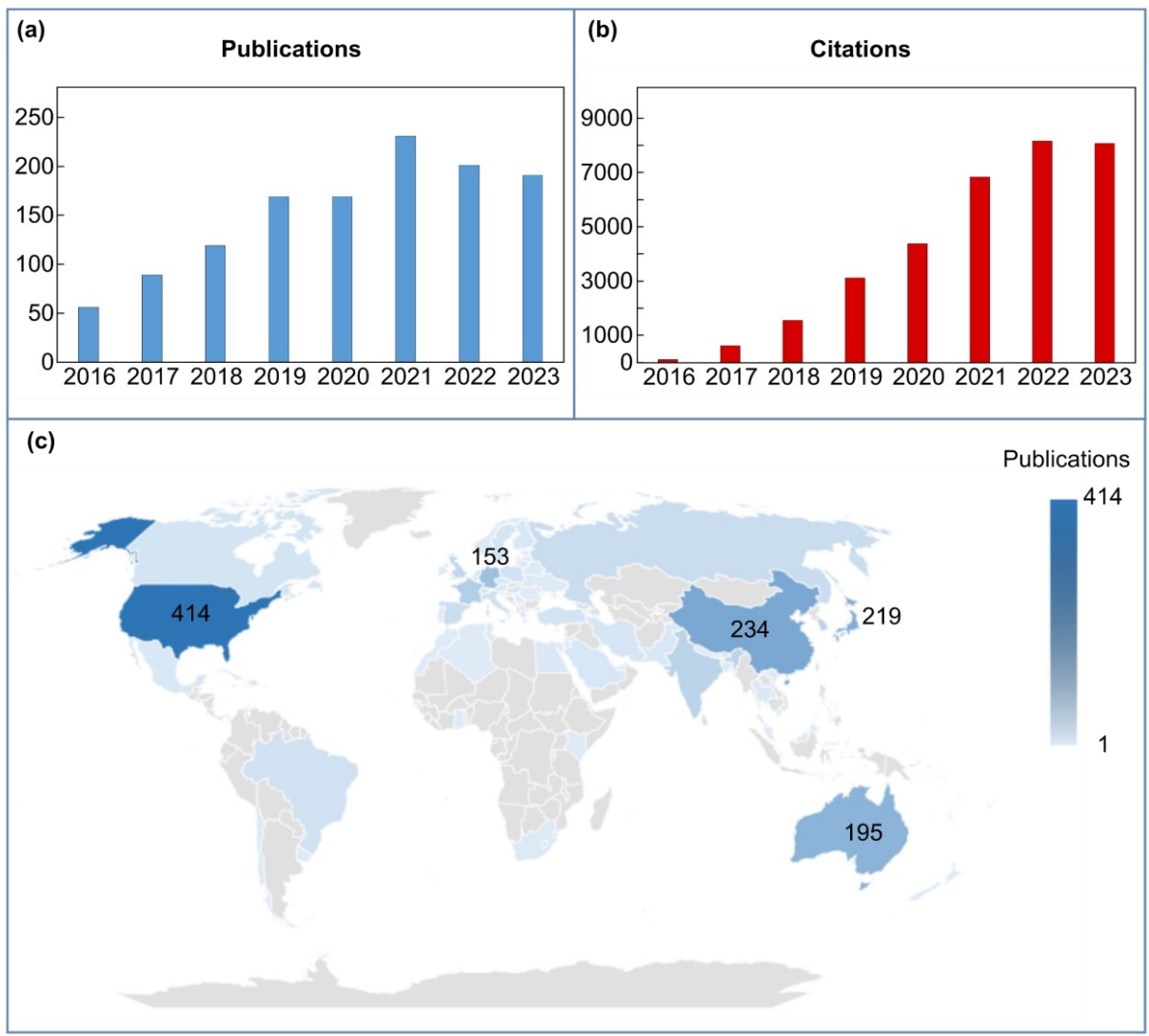

**Figure 2: Statistics on global scientific research concerning hBN quantum emitters from 2016 to 2023. (a, b)** Number of publications and citations using the Web of Science (Clarivate Analytics) with the keywords ["hexagonal boron nitride" OR "hexagonal boron-nitride" OR "hBN" OR "h-BN" (Topic) and "quantum emitter*" OR "color cent*" OR "color-cent*" OR "quantum light*" OR "single-photon*" OR "single photon*" OR "emitter*" OR "ensemble*" OR "dipole*" OR "quantum-emitter*" OR "quantum-light*" OR "defect*" (Topic) and "fluorescence*" OR "emission*" OR "photon*" OR "light*" OR "optical"]. **(c)** Corresponding distribution of publications by countries, powered by Bing © Australian Bureau of Statistics, GeoNames, Microsoft, Navinfo, Open Places, OpenStreetMap, TomTom, Zenrin. The search was performed on August 09, 2024.

## 4. Theoretical and experimental investigations on the origin of various defect families

Unlike the case of most color centers in diamond, the exact chemical structures of most quantum emitters in hBN are largely unknown. This is in part due to the complexity involved within the hBN lattice. While diamond contains only a single element, carbon, hexagonal boron nitride comprises two elements, boron and nitrogen. Such poly atomic nature results in more possible defect configurations in hBN than in diamond. Additionally, hBN has more lattice imperfections and impurities than diamond, rendering the determination of the responsible defect centers a daunting task. Another factor is the built-in strain in hexagonal boron nitride. As hBN is a two-dimensional material, its single-layer sheet of atoms can sustain unusually high lattice strain of up to 5%, similar to graphite and other transition metal dichalcogenides (TMDs). Based on theoretical calculations, the built-in strain can shift the ZPL of an emitter up to tens or hundreds of millielectronvolts.[25] Such a phenomenon significantly complicates the assignment of the defect structures since it can be mistakenly grouped into another defect family whose ZPL energy is close by.

Quantum emitters in hBN can be categorized into four families according to their emission wavelengths: UV (~4.1 eV), blue (~2.85 eV), (~2 eV) visible, and spin-active (~1.5 eV) defects.[3] Variations in the emission wavelengths within each group are likely attributed to local built-in strains. In the UV-emitter group, carbon-related defects, such as the substitutional carbon at the nitrogen site ($C_N$) and the carbon dimers ($C_B C_N$) are proposed to be responsible for the emission.[26-27] The so-called blue-emitters have recently been found and tentatively attributed to the carbon split interstitial defect ($C_N^2$).[28] The visible emitters are the most studied among the four families thanks to their extraordinary optical properties, including ultrahigh brightness, linear polarization, and controllable emission wavelengths via external stimuli such as strain or electric fields. Emitters in this family cover a wide range of emission wavelengths ~550–850 nm (2.25–1.46 eV). Initially, most of these visible emitters are attributed to non-carbon-related configurations such as the anti-site nitrogen-vacancy ($N_B V_N$).[12] Carbon is later thought to be involved in the chemical structure of these defects, such as the carbon antisite $C_B V_N$. Some defects in this group have recently been found to be spin-active and can be initialized, manipulated, and read out all optically—a technique called optical detected magnetic resonance (ODMR). Recent work suggests that the negatively charged $V_B C_N^-$ can be a candidate responsible for such optical characteristics.[22-23] Overall, the most popular theoretical framework used thus far is the density functional theory (DFT), thanks to its efficiency, versatility, and accuracy for ground-state calculations.[29-31] DFT calculations are being used as a powerful tool in providing insightful understanding into the electronic structure of the materials and guiding the design of catalysts and functional materials with superior properties.[32-39] However, the use of conventional level of theory is not accurate in DFT studies on h-BN material, particularly when evaluating defect states, and therefore, higher level of theory is required.[5, 40-41] Reimers et al. comprehensively calibrated the performance of different DFT methods in describing spectroscopic and energetic properties of h-BN defect sites.[40] This study reported that DFT methods using the conventional generalized-gradient approximation functionals such as PBE performed poorly and could not be applied to defect states and higher levels such as hybrid functional HSE06 and long-range corrected functional CAM-B3LYP. Although those higher levels of theory implement more computational cost, they are making significant progress in shedding light on the origin of various defects sites in h-BN.

Huang et al.[42] used DFT calculations to gain mechanistic understanding on the formation of defects in hBN at microscopic scale and revealed the crucial role of interdefect electron pairing in stabilizing the donor- and acceptor-type defects combination. Dorn et al. used the integration between DFT calculations and high-resolution solid-state NMR spectroscopy to identify the detailed structure of the different defect sites within h-BN.[43] Besides providing structural information, DFT calculations also delivered insightful information on electronic properties of the defect sites. The stable spin states, charge transition levels, and optical excitations were computed at various defect centers in h-BN, including boron antisite, boron vacancy, nitrogen vacancy, and nitrogen antisite; and the $C_BV_N$ defect was identified as the highly potential candidate for application in quantum bit and emitting.[44-46] The unprecedented strong coupling of defect emission in hBN to stacking sequences was reported in Li et al.[47] shedding light on the design principles for precise controlling of defect emission. The optical properties of boron vacancy defect center were also evaluated in Ivady et al.[48] revealing the corresponding routes responsible for the spin dependent luminescence and optical spin polarization of this defect center. Recently, a notable effort in this area has resulted in a comprehensive online database of hBN optically active defect centers,[49] covering 257 triplet and 211 singlet electronic structures. In this work, the Vienna Ab initio Simulation Package (VASP) is used in conjunction with the HSE06 exchange-correlation functional to efficiently screen through many defect configurations. At first, 158 defect complexes from group III–V are considered. Subsequent calculations determine the spin multiplicity: singlet, doublet, or triplet. Additional charges (positive or negative) are added for the doublet cases, followed by the geometrical optimizations. Some of the details are shown in **Figure 3a**. The defect complex is then added to the online library (https://h-bn.info). Interestingly, many useful parameters of these defects are collected, including their ZPL position, photoluminescence spectra, excitation and emission polarization, radiative lifetime, and quantum efficiency—making it among the most extensive theoretical investigations on hBN quantum emitters to date. Furthermore, with the recent development of machine learning techniques,[50-51] DFT calculations are expected to have more important role in advancing the knowledge of quantum bits in hBN-based materials.[52-53]

However, it has been shown that in some cases, a rigorous experimental framework alone could pinpoint a particular defect configuration with high confidence. Such an example is the discovery of the so-called spin-active defect family, which has an emission wavelength of ~850 nm. In this case, electron spin resonance angle-dependent (EPR) and ODMR, together, unveil the atomic origin of the defect, the negatively charged boron vacancy ($V_B^-$), as depicted in **Figure 3b**. Similar to the third defect family (the carbon-related spin-active group), the $V_B^-$ defect complex has a spin multiplicity of $S = 1$, meaning that it has a spin triplet with $m_S = \pm 1$ and $m_S = 0$ manifolds, with a zero-field splitting of ~3.5 GHz. Without the help of DFT, such a robust experimental framework has proven to be very effective in unveiling the chemical structure of these defect complexes in hBN. However, this situation is not always applicable in all cases especially with defect complexes comprising multiple impurities and vacancies. It is, therefore, challenging to universally implement the scheme to all observed defect complexes. Recent efforts have been spent on correlative studies between optical and electron microscopy. For instance, a combination of photoluminescence, cathodoluminescence, and electron diffraction was attempted to gain insights into the atomic structures of the visible defect family.[54] While the joint approach was not successful in pinpointing the origin of these defects, it showed that a cross-platform correlative method is highly valuable in extracting hidden information about these defect complexes.

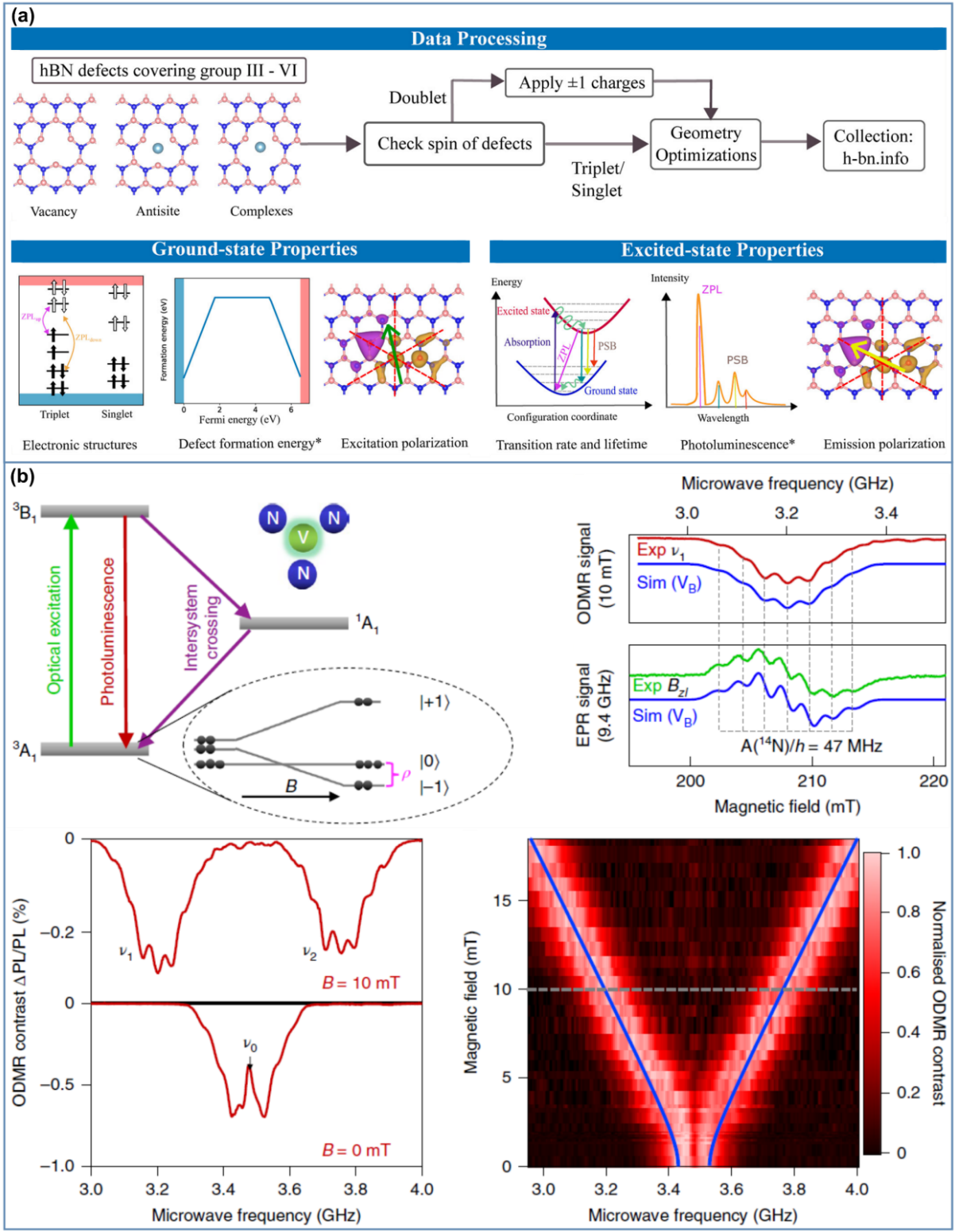

**Figure 3: Theoretical calculations and experimental research on the origin of hBN quantum emitters. (a)** The hBN defects database: A Theoretical compilation of color centers in hexagonal boron nitride **(b)** hBN spin defects read-out by electron paramagnetic resonance spectroscopy and optically detected magnetic resonance measurements at room temperature. Figure a adapted under terms of the CC-BY 4.0 license.[49] Copyright 2024, The Authors,

published by American Chemical Society. Figure b adapted with permission.[24] Copyright 2020, Springer Nature.

## 5. Host material fabrication and defect engineering

Compared to other host materials such as diamond or silicon carbide, hexagonal boron nitride bulk crystals and powder are significantly cheaper. Hexagonal boron nitride host can be categorized into five groups: chemical vapor deposition (CVD), metal–organic vapor-phase epitaxy (MOVPE), solvent-exfoliated nanoflakes, mechanically-exfoliated microflakes, and high-pressure high-temperature (HPHT) bulk crystals as shown in **Figure 4a**. Chemical vapor deposition is a well-established method for growing hBN thin film with controllable thickness.[55-57] The technique involves the controlled flow of gaseous precursor(s) into a quartz tube heated at temperatures exceeding the decomposition threshold of the precursor(s). As such, the technique is straightforward and requires inexpensive apparatus. One of the most significant advantages of this method is its scalability. Previous studies showed CVD growth at wafer-scale sizes,[58-59] enabling prospects of industrial integration. By adjusting the growth parameters such as temperature, duration, gas partial pressure, and precursor types, optically-active defect centers were embedded in situ.[60-61] The emission spectra from these CVD-grown hBN films feature a wider full-width-at-half-maximum (FWHM) and lower peak intensity than those from solvent-exfoliated nanoflakes. In addition, quantum emitters in these films tend to be susceptible to blinking and bleaching. The underlying reasons for this phenomenon remain largely unknown. Although there are still no techniques to effectively tune the defect density in these CVD hBN films, growth on pillar-patterned substrates exhibited a certain degree of control over the defect position within the films.[62] The following method is MOVPE, in which triethylboron and ammonia are the precursors. The technique is significantly more complex than CVD. Most notably, the technique demands highly pure metalorganic compounds and hydrides, and its operating parameters need to be controlled and synchronized precisely. The synthesis technique witnessed carbon incorporation during the hBN growth, resulting in thin hBN film with various densities of quantum emitters, depending on the input carbon concentrations within the precursors. While this method has only been introduced recently, it has gained significant traction due to its ability to produce optically-addressable spin-active defect centers.[22-23] As such, MOVPE is expected to receive increasing attention in the research community even though the technique requires significant facility investment owing to its complex and stringent operations. The third method is solvent exfoliation. This technique is the least expensive since it relies on an inexpensive starting material, hexagonal boron nitride powder. Hexagonal boron nitride powder is typically synthesized via ammonolysis of boric oxide at high temperatures, from $800 - 1200°C$.[63] The obtained powder is dispersed in suitable solvents such as ethanol, isopropanol (IPA) or N-methylpyrrolidone (NMP), whose surface energies match the cross-plane interactions between hBN sheets. The hBN flakes are gradually exfoliated in the solution by the sheering motions induced by the ultrasonication bath. The cavitation effect creates randomly distributed, localized spots with extremely high temperature and pressure, causing the flakes to separate in the out-of-plane direction.[64] Prolonged sonication promotes thinner hBN flakes in the solution at the expense of smaller flake diameters due to the breakdown of the flakes due to the sonication process. The resultant solvent-exfoliated flakes possess a typical dimension of 200–500 nm in diameter and 20–100 nm in thickness.[64] The defects are embedded within the exfoliated flakes and can be activated by simple annealing at a temperature $> 850°C$.[25] Although inferior in terms of crystallinity compared to CVD and MOVPE, solvent exfoliation provides a low-cost alternative for applications where high-quality crystalline material is not required, such as heterogeneous integration of the quantum emitters with on-chip photonic architectures or quantum key distribution. The fourth approach is bulk hBN grown by high-pressure high-temperature (HPHT) process, where temperature and pressure often exceed 1500°C and 4.5 GPa, respectively.[65] This approach yields the purest hBN crystals among the fabrication methods mentioned above. Most notably, bulk hBN fabricated by Taniguchi, Watanabe, and co-workers at the National Institute for Materials Science (NIMS) Japan, is widely regarded as the highest quality hBN to date and used for heterostructures in two-dimensional devices around the world.[66] As such, the material rarely contains optically active defects, and therefore, defect engineering techniques are required to implant suitable defects into the hBN lattice. Many native quantum emitters in bulk hBN exhibited significant optical blinkings and possessed multiple dark manifolds,[12, 67] making them unsuitable for most quantum applications. The final form of hBN host is the mechanically exfoliated flakes, which are exfoliated by the Scotch tape method.[68] Though simple, the method can produce high-quality flakes with large sizes. The typical flakes feature a thickness ranging from a few layers to a few hundred nanometers and a lateral dimension of a few to tens of microns. On the one hand, the technique is often used in the laboratory setting owing to its lack of scalability. On the other, the most important advantage of this method is its simplicity and versatility, which allows for rapid building of complex 2D heterostructures from various 2D materials whose lattice parameters are mismatched. For these reasons, mechanically exfoliated hBN is a preferred choice for device fabrication, such as electrically driven single-photon emitting devices or widely tunable single-photon light sources.[69-70]

We now turn to defect engineering strategies for the hBN hosts (**Figure 4b**). The first method involves thermal annealing at high temperatures, typically > 850°C. At these temperatures, quantum emitters were observed to be activated, and the quantity seemed to be weakly correlated with the annealed temperature.[25] However, the exact mechanism for such activation remains elusive. The thermal treatments effectively activate optically active defects in various hosts, including bulk, solvent- and mechanical-exfoliated flakes.[64, 71-72] In addition, the method is also used as a post-treatment for a main structural modification process, as discussed later in the text. Though simple and effective, thermal treatments lack spatial determinism and are challenging to implement on substrates/structures that are temperature-sensitive. The second approach, laser writing, for example, tackles this issue. In this approach, a femtosecond laser (pulse width <1000 fs) is tightly focused onto an hBN flake, releasing pulse energies in the order of tens of nanojoules.[18] Beyond a certain pulse energy threshold, laser-induced damage was observed, with defective structures ranging from bubble-like to crater-like formations. Bright and stable quantum emitters were observed after the subsequent thermal annealing treatment. The yield of such a technique reached ~43%, which is relatively high among other competing methods. Nevertheless, the mechanism of defect formation is still poorly understood, making it difficult for further improvements to occur. Plasma-induced etching is another promising technique for creating emitters in hexagonal boron nitride. This method relies on the mild etching created by the accelerated ions in the plasma.[17, 73] The ion species can be selected by introducing different types of gas, such as oxygen, argon, and hydrogen.[17, 72-73] These highly energetic ions bombard the hBN lattice, causing bond breakage and vacancies, and the defect density is shown to have a strong correlation to the plasma power. The subsequent thermal treatment then induces lattice reconstruction, which promotes the formation of optically active defect centers. When executed with a lithography-defined mask, the plasma etching process can create an array of site-specific quantum emitters. The following method is electron-beam-induced defect creation. In this technique, highly accelerated electrons in a scanning electron microscope (SEM) are directed toward hBN flakes, inducing atomic-sized defects in the lattice. Due to the negatively charged nature of the electrons, their interactions with the lattice are complicated and require simulation platforms such as Monte Carlo simulation of electron trajectories (CASINO) in solids to extract crucial interaction parameters.[74] These parameters include electron trajectories, energy loss, backscattered, and secondary electron generation. Early attempts using electron beams to create emitters in hBN resulted in randomly distributed optically active defects that tend to localize near edges and grain boundaries.[25, 75] Recent studies, however, showed that electron irradiation effectively created quantum emitters in a deterministic manner. The method produced emitters from both the visible and blue families.[76-78] Unlike the visible family, the blue emitters possess a narrow linewidth distribution of ~3 meV,[77] allowing for the first two-photon quantum interference from emitters in hBN.[79] A closely related technique is ion implantation which arises from an ion implanter or a focused ion beam (FIB). Previous studies reported increases in emitter density with various ion species.[75] The spatial distribution of optically active defects was, however, arbitrary. Spatial localization of the quantum emitters was recently achieved by gallium FIB,[80] with a yield as high as 31%. The milling process created an array of holes with edges at their circumferences, where hBN emitters were located. Further optimization of the ion implantation conditions can be obtained by detailed simulations on platforms such as the stopping range of ions in matter (SRIM) since it can provide insightful information about ion range and distribution, energy loss, sputtering yield, and vacancy concentration.[81] Another method involves using indentation by an atomic force microscope (AFM) to induce "crater-like" pits on the hBN flakes.[82] The damage created at the rim of the pits was found to host quantum emitters, with a yield of ~36%. The ZPL distribution of the emitters fabricated by this technique is relatively wide, however, ranging from ~582–633 nm, and the PL lineshapes vary significantly from emitter to emitter. Another drawback of the method is its destructive nature—making it difficult to seamlessly integrate into on-chip photonic structures.

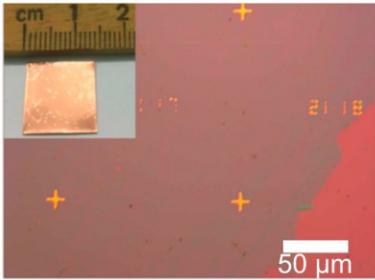

(a) hBN structures: CVD, MOVPE, Solvent exfoliation, HPHT, Mechanical exfoliation. (b) Defects engineering: Thermal annealing, Laser writing, Plasma, Electron beam irradiation, Ion implantation, Nanoindentation.

**Figure 4: Fabrication of hBN quantum emitters. (a)** Five general methods to create hBN structures: chemical vapor deposition (CVD), metal-organic vapor-phase epitaxy (MOVPE), solvent-exfoliated nanoflakes, high-pressure high-temperature (HPHT) bulk crystals, mechanical-exfoliated microflakes. **(b)** Schematic representations of six methods to create optically active defects in hBN: thermal annealing, laser writing, plasma, electron beam irradiation, ion implantation, and nanoindentation. Figure optical image CVD hBN adapted with permission.[60] Copyright 2019, American Chemical Society. Figure SEM image MOVPE hBN adapted under terms of the CC-BY 4.0 license.[83] Copyright 2021, The Authors, published by Springer Nature. Figure SEM image solvent exfoliation hBN adapted with permission.[64] Copyright 2021, American Chemical Society. Figure optical image HPHT hBN adapted with permission.[71] Copyright 2016, American Physical Society. Figure optical image mechanical exfoliation hBN adapted with permission.[17] Copyright 2018, American Chemical Society.

## 6. Monolithic integration into photonic cavities and waveguides

Most quantum applications demand the quantum light source to be emitted into a particular optical mode that is coherent and directional. Therefore, integrating individual hBN quantum emitters into photonic cavities and waveguides is a vital step toward any practical implementations. Such incorporation schemes can be classified into two main categories: monolithic and non-monolithic integrations. In the monolithic scheme, the emitter and the photonic structures share the same host material—hexagonal boron nitride. It must be noted that hexagonal boron nitride is a dielectric that has a birefringence value $\Delta n = n_\perp - n_\parallel = \sim -0.35$,[11] which needs to be taken into account in the design. The main advantage of this approach is that the photon coupling from the emitter to the cavity or waveguide is usually superior to that of the non-monolithic counterpart. This is because the emitter can be accurately positioned at the optimal location within the photonic structures, rendering maximum coupling efficiency. The fabrication process typically involves several steps. First, a type of photonic structure, say a cavity, is designed using finite-difference time-domain (FDTD) software where the resonant wavelength and the quality (Q) factor can be calculated. Second, the fabrication of the cavity structure follows the standard electron beam lithography (EBL) process on hBN flakes deposited on silicon substrates. The procedure entails e-beam resist patterning, developing, dry-etching (sulfur hexafluoride, $SF_6$), undercut (KOH solution), and thermal annealing. Third, the emitters are created by a site-specific method, such as electron beam irradiation, to produce individual emitters at locations with maximal field intensities. Using this technique, one-dimensional (1D) photonic crystal cavity (PCC) was created, and emitter-cavity weak-coupling was demonstrated,[84-85] as shown in **Figure 5a**. The estimated Purcell factor—a photoluminescence enhancement ratio of the coupled emitter to the bare emitter—was estimated to be ~15. Such a value is considerable, considering that dielectric cavities often feature significantly larger mode volumes compared to their plasmonic counterparts.

By further fine-tuning the fabrication procedure by employing a combination of low $SF_6$ concentration, low gas partial pressure, and low etching power, Nonahal and co-workers showcased a slow etching process, resulting in very smooth sidewalls in the photonic structures. The smoothness of the structures gives rise to Q factors of the two-dimensional (2D) PCC as high as 2000 (**Figure 5b**). Microdisk resonators were fabricated using the same approach, exhibiting Q factors exceeding 3000 (**Figure 5c**). Complicated photonic structures such as the bound state in the continuum (BIC) metasurfaces have also been fabricated monolithically.[86] Coupling of spin-active defects to the structures were reported, yielding 25-fold increase in photoluminescence and spectral linewidth reduction to under 4 nm. Couplings of quantum emission into a waveguide mode were demonstrated by Li et al. using a similar technique, which resulted in a relatively modest coupling efficiency of ~3% (**Figure 5d**).[87] The monolithic fabrication of hBN photonic architectures has advanced significantly during the last several years, with suspended, fully integrated quantum photonics (IQP) recently realized.[88] Continuous improvements in the fabrication steps, especially dry-etching and mask removal, are needed to reach the Q factors closer to what was achievable in bulk dielectrics such as silicon and III-V semiconductors. Ultimately, the hard limit in the performance might be set by the two-dimensional nature of the material per se. The weak out-of-plane interactions between atomically thin sheets of hBN make it challenging to sustain a high degree of smoothness during the etching process, unlike other bulk materials, such as silicon, whose lattice integrity is isotropic.

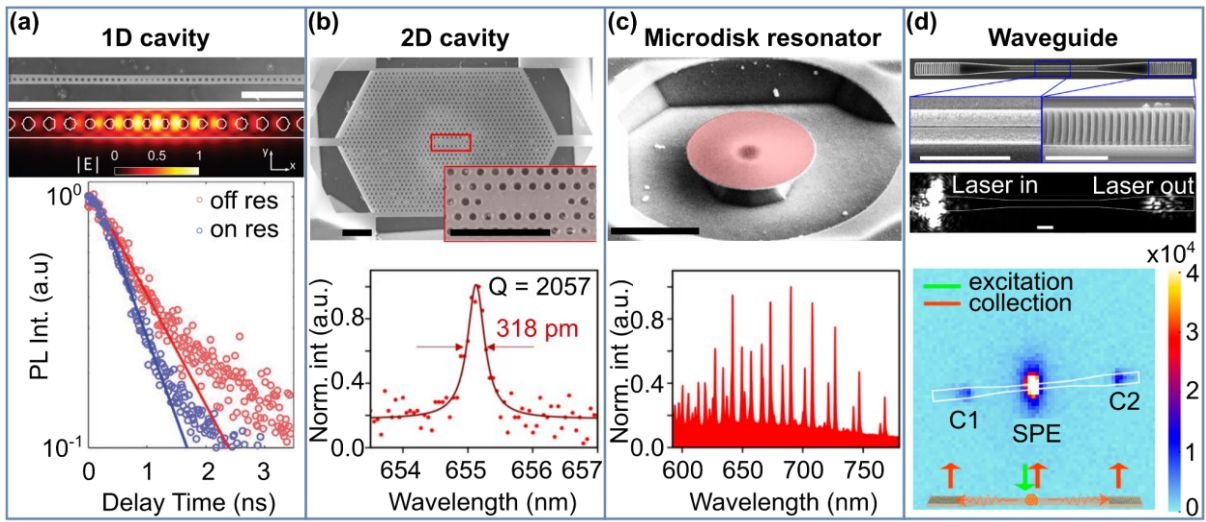

**Figure 5**: Monolithic integration of hBN quantum emitter in photonic structures. (a-d) SEM images and representative optical signatures of 1D, 2D photonic crystal cavities, microdisk resonator, and waveguide, respectively. Scale bars in all figures are 2 $\mu m$. Figure a adapted with permission.[84] Copyright 2021, Wiley-VCH GmbH. Figure b, c adapted with permission.[88] Copyright 2023, Wiley-VCH GmbH. Figure d adapted with permission.[87] Copyright 2021, American Chemical Society.

## 7. Hybrid integration into dielectric and plasmonic architectures

In the non-monolithic category, two platforms are considered: hybrid dielectric and plasmonic integration. The form scheme encompasses the positioning of individual quantum emitters embedded in hBN micro flakes or nanoflakes atop pre-fabricated dielectric structures. Compared to the monolithic counterpart, the hybrid approach is significantly more versatile since various dielectrics, such as silicon, silicon carbide, or III–V semiconductors, can be used as the material for fabricating cavities or waveguides. As such, the approach makes integrating hBN quantum emitters into well-established integrated photonic chips a straightforward task. Hybrid dielectric architectures, however, suffer from a significant weakness—that is, the suboptimal coupling efficiency of the emitter to the photonic structures. The quantum emitters are usually placed on the dielectric cavities or waveguides. At the same time, the maximal field intensities are located inside the photonic formations, resulting in a spatial mismatch between the two. An example of this strategy was conducted by Froch et al. Here, an hBN film grown by CVD was transferred onto a silicon nitride (SiN) substrate, and the composite was patterned with EBL and dry-etched into hybrid CVD hBN/1D PCC structures, as depicted in **Figure 6a**. A quality factor of 2400 and a PL enhancement of 9 folds were reported.

Vogl and co-workers integrated quantum emitters in mechanically exfoliated microflakes into microcavities (**Figure 6b**). The cavities are a modified Fabry-Perot type consisting of a hemispherical and a flat mirror that confines the cavity mode to the emitter location. The cavity mode is tunable thanks to the adjustable thickness of the sandwiched PDMS layer between the two mirrors. When on resonance, the cavity-emitter composite features a Purcell factor of 2.3 and a 25-fold reduction in linewidth values. Owing to the compact design of the microcavities and other miniaturized optical components, the single-photon source setup is entirely self-contained within an *10 × 10 × 10 cm³* enclosure. Unlike the previous two designs, the third example demonstrates a coupling of hBN quantum emitters, the $V_B^-$ defect center in particular, to the plasmonic nanopatch antennas (NPAs),[89] as shown in **Figure 6c**. The hybrid system comprises a bottom thin silver film, a thin hBN flakes, and a top silver nanocube. Such a design not only offers extremely tight confinement of electromagnetic fields of a plasmonic cavity but also produces directional emission of a plasmonic antenna. As a result, an overall PL enhancement of ~250 folds was achieved, significantly better than any other coupling schemes.[90-91] A drawback of this design, however, lies in the probabilistic nature of the coupling since the extreme mode confinement induced by the silver film and nanocube requires the perfect spatial matching of the emitter to the mode. Besides cavities, hybrid waveguides were also demonstrated using a similar strategy. For instance, Kim and co-workers fabricated the hybrid hBN-AlN composite waveguide structures by placing the hBN nanoflakes on top of the pre-existing aluminum nitride formations (**Figure 6d**).[92] The significant spatial mismatch between the emitter and the waveguide mode and the mode leakage induced by the underlying sapphire substrate resulted in a modest coupling efficiency of < 2%. Other more sophisticated integrations between hBN emitters and photonic structures such as metalenses or BICs were also realized experimentally.[93-94] The former showcased the ability to spatially separate quantum emission with different polarization states. The latter featured the first room-temperature strong-coupling between an hBN quantum emitters and a cavity, resulting in a noticeable Rabi splitting of ~4 meV. Though challenging, the spatial mode mismatch issue mentioned above can be tackled. An innovative strategy has recently been demonstrated using a composite SiN-hBN microring resonator structure in which the hBN flake was sandwiched between the top and bottom halves of the SiN resonators.[95] Such a composite structure addressed the spatial matching issue mentioned above, though with a significantly higher degree of complexity in the fabrication process. Such a clever integration is a step in the right direction to improve the performance of the hybrid photonic architectures, rivaling those of monolithic counterparts.

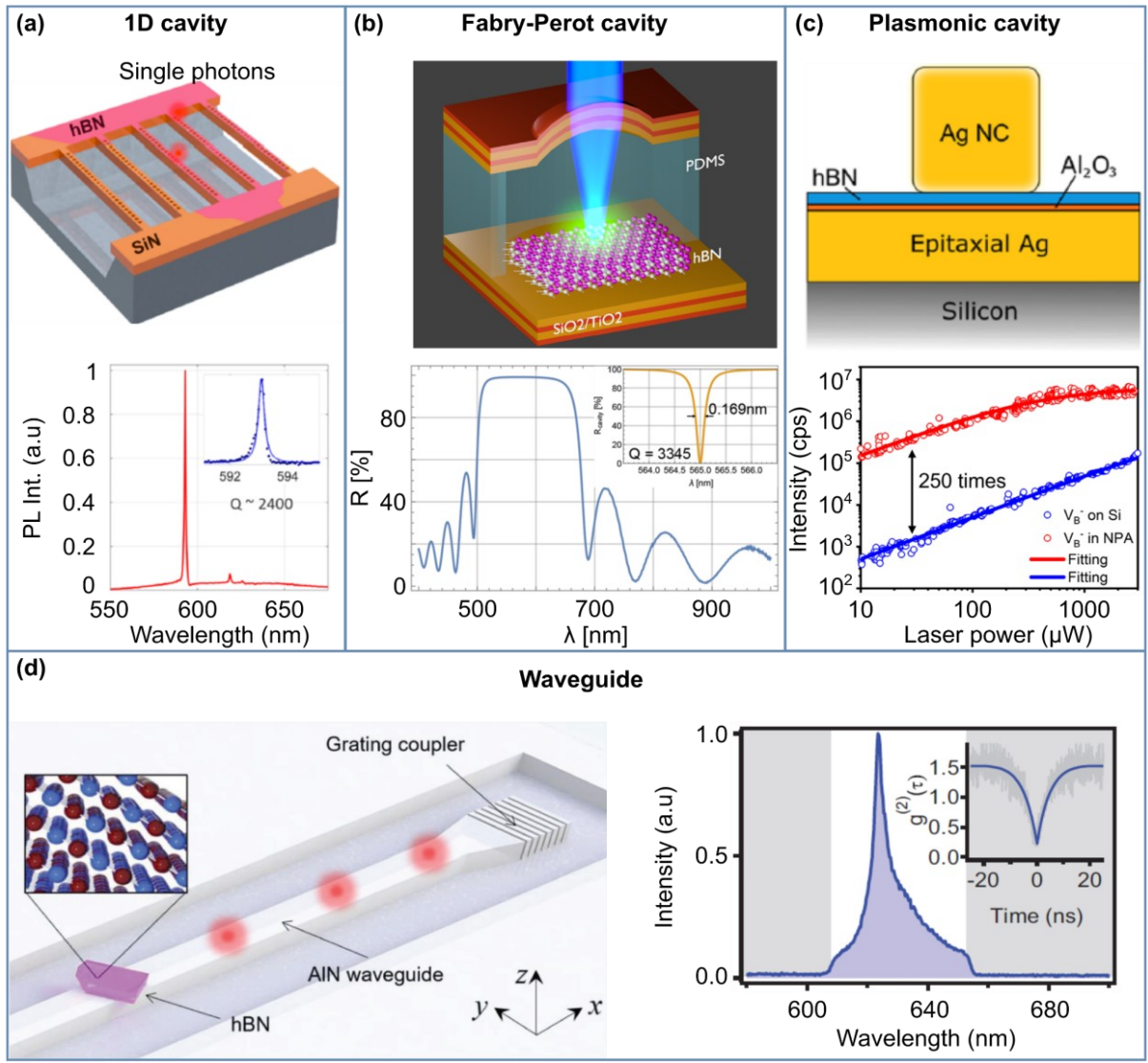

**Figure 6:** Non-monolithic integration of hBN quantum emitter in photonic structures. (a-d) Schematic and corresponding results of various hybrid integration of hBN quantum emitters: one-dimensional photonic crystal cavity, Fabry-Perot cavity, plasmonic cavity, and waveguide structure, respectively. Figure a adapted with permission.[96] Copyright 2020, American Chemical Society. Figure b adapted with permission.[97] Copyright 2019, American Chemical Society. Figure c adapted with permission.[89] Copyright 2022, American Chemical Society. Figure d adapted with permission.[92] Copyright 2019, Wiley-VCH Verlag GmbH & Co. KGaA, Weinheim.

## 8. Applications in quantum sensing

Quantum emitters in hBN have recently emerged as a promising alternative to established platforms such as color centers in diamonds or silicon carbide. Hosted in atomically thin sheets of hexagonal boron nitride, quantum emitters can be positioned closer to the material surface without being severely disturbed by the surface states and trapped charges, thanks to the dangling-bond-free nature of the material. One of the simplest examples of quantum sensing is quantum thermometry—the measurement of local temperature using quantum systems. Chen et al. reported using individual quantum emitters in hBN nanoflakes to monitor temperature down to the sub-micron scale (**Figure 7a**).[98] The spectral characteristics of hBN quantum emitters, such as the broadening of the linewidths or the shifting of the ZPLs, are well-correlated with the temperature changes. As a result, local temperature readouts at different spots of a microheating circuit were demonstrated by an individual quantum emitter. Since the emitter is around a nanometer in size, the work implies that a few-nanometer-sized thermal probe can be realized with further improvements in the material engineering aspect. Another exciting example is nanofluidic sensing, reported by Radenovic and co-workers (**Figure 7b**). In their work, surface defects in hBN were submerged inside solvents with various polarities, such as ethanol, acetonitrile, dimethylsulfoxide, pentane, etc. Owing to the dipole-dipole interaction between the defects and the solvent molecules, the defects became optically activated and emitted photons. Solvents with stronger dielectric induced stronger red-shift of the emission from the quantum emitters due to the enhanced dipolar interactions. Levering such a mechanism, hBN quantum emitters were used to image confined solvents down to the nanoscale and probe the changes in the liquid dielectric constants in nanoscale-confine circumstances. The work demonstrated the potential use of hBN quantum emitters for studying the liquid-solid interface at the nanoscale. Quantum emitters in hexagonal boron nitride were also utilized for magnetic and temperature sensing (**Figure 7c**). In particular, Healey et al. fabricated thin hBN flakes hosting the negatively charged boron vacancy defects by ion implantation. The defects are optically active and can be initiated, manipulated, and read out using the ODMR technique. By leveraging the even distribution of these near-surface quantum sensors, the team showcased the multimodalities of the technique by simultaneously mapping thermal and magnetic field distributions in a 2D magnet $CrTe_2$.[99] They also demonstrated visualizing the electron currents and Joule heating in a working graphene-based device. Measuring other physical parameters, such as pressure and strain, was also realized using the $V_B^-$ defects, from other research groups.[100-101]

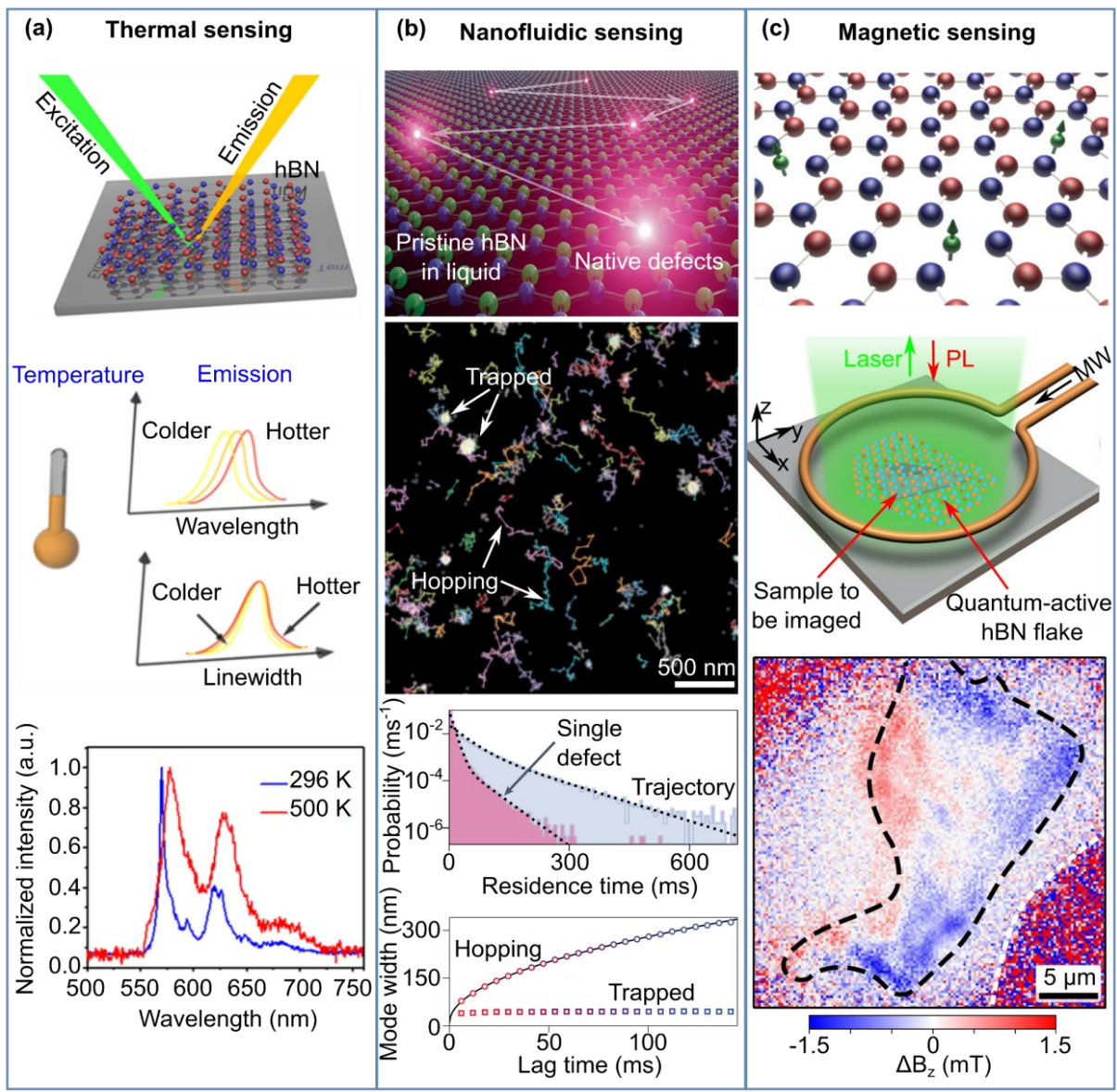

**Figure 7: Application of hBN quantum emitters in quantum sensing. (a)** Optical thermometry with quantum emitters in hBN nanoflakes **(b)** Nanofluidic sensing with quantum emission from pristine hBN **(c)** Quantum microscopy with spin defects in hBN. Figure a adapted with permission.[98] Copyright 2020, American Chemical Society. Figure b adapted under terms of the CC-BY 4.0 license.[102] Copyright 2023, The Authors, published by Springer Nature. Figure c adapted with permission.[24] Copyright 2020, Springer Nature.[99] Copyright 2023, Springer Nature.

## 9. Applications in quantum cryptography

Besides quantum sensing, quantum cryptography—quantum key distribution (QKD)—is another area where hBN quantum emitters are expected to shine. Quantum key distribution has gained tremendous traction recently, owing to its ultrasecure encryption procedure governed by the no-cloning theorem in quantum mechanics. The quantum emitters are ultrabright, room-temperature operational, feature high single-photon purity, and possess linearly polarized emission, the criteria demanded in QKD. Based on the BB84 protocol, Al-Juboori et al. demonstrated quantum key distribution in free-space using quantum emission (ZPL at 645 nm) from an hBN emitter.[103] In their experiment protocol (**Figure 8a**), single photons with randomly selected polarizations from the four possible states, horizontal (H), vertical (V), right-handed (R) and left-handed (L) circular, were synchronized with the exciting pulsed laser and prepared by the first electro-optic modulator (EOM) at Alice, the sender. These polarized photons were sent to Bob, the receiver, and measured by a pair of SPADPs using one of the two bases, H/V or R/L, thanks to the second EOM and the time-gated detection based on the pulsed laser. Once finished, Alice and Bob exchanged the bases used for each photon via a public channel, but they did not share the measurement results. Together, they discarded photons measured by the wrong bases and kept the correct ones. With additional error correction algorithms and privacy amplification, the string of the correct photons formed the secret key. A secret key of ~70000 bits and a security level of $10^{-10}$ were reported. Quantum key distribution was also demonstrated in a water environment by the same team using a blue emitter whose ZPL is at ~ 436 nm (**Figure 8b**).[104] It must be noted that the team only simulated the underwater transmission conditions by using water tubes between the optical components. Such fluorescent emission is desirable since it closely matches the dip in the water absorption spectrum (~ 417 nm).[105] Although the underwater performance was inferior to that in the air, the study laid a foundation for further developments in water-based quantum key distribution. Unlike the underwater setting, QKD during daylight conditions requires a different transmission wavelength. Vogl and co-workers identified the H$\alpha$ Fraunhofer line at 656 nm as the optimal wavelength for QKD protocols in such conditions.[106]

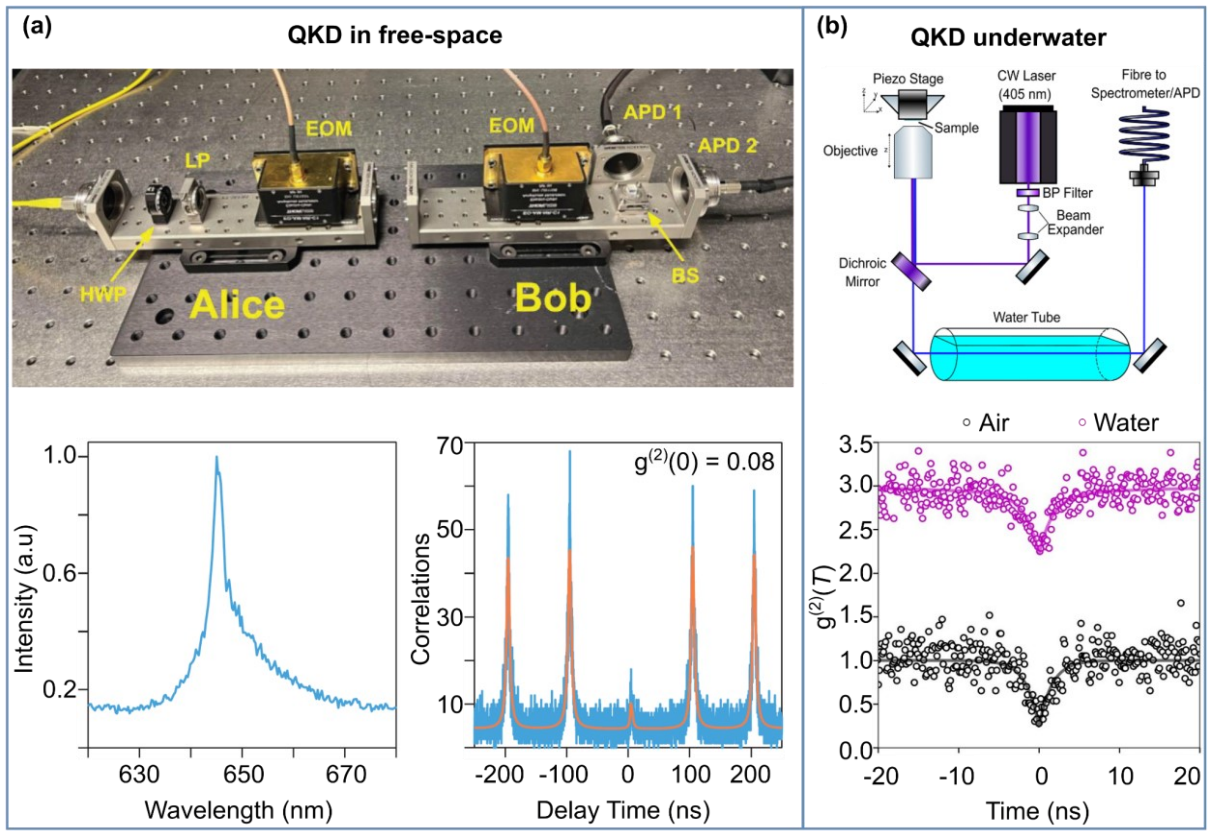

**Figure 8: Application of hBN quantum emitters in quantum key distribution (QKD). (a)** The first proof of finite-key BB84 QKD system realized with hBN quantum emitters in free-space (optical components of the transmitter (Alice) and the receiver (Bob) setup including EOM: electro-optic modulator, LP: linear polariser, APD: avalanche photodiode, PBS: polarizing beam splitter, HWP: half-wave plate **(b)** Optical communications with hBN quantum light sources underwater. Figure a adapted with permission.[103] Copyright 2023, Wiley-VCH GmbH. Figure b adapted under terms of the CC-BY 4.0 license.[104] Copyright 2024, The Authors, published by IOP Publishing Ltd.

## 10. Applications in quantum communication

Another critical area where hBN quantum emitters can be deployed is quantum communication. A prerequisite for such applications is the ability to produce indistinguishable photons—photons that can interfere with each other. For a light source to generate indistinguishable photons, it needs to (i) have high single-photon purity, (ii) produce photons with the same wavelength, polarization, and spatial mode, and (iii) create photons whose spectra are Fourier-transformed-limited—their linewidths defined only by the excited state lifetime.[107] Fournier and co-workers established the first experimental demonstration of the two-photon quantum interference experiment, the Hong-Ou-Mandel (HOM), using a blue emitter in hBN.[79] In their experiment (**Figure 9a**), the emitter was excited non-resonantly with an 80MHz pulsed laser, and the emission from the ZPL was exclusively collected using a combination of high-resolution grating and single-mode-fiber coupling. By introducing a 12.5-ns delay time, the two photons interfered with one another on a fiber-based beam-splitter, resulting in a HOM dip of 0.32 for the parallel polarization case, corresponding to an indistinguishability of 0.56. The deviation from a perfect HOM dip (zero) was due to the dephasing caused by the non-resonant excitation. It was calculated that the indistinguishability could approach 90% if the emitter were coupled to a cavity (Purcell factor of ~15) and excited resonantly. The work shed light on using blue emitter in hBN for quantum applications that require HOM protocol. In another attempt, Vogl and co-workers proposed a quantum photonic circuit to test an extended theory in space using hBN quantum emitters as a single-photon source as shown in **Figure 9b**, top panel.[108] Specifically, the team conceptualized a circuit that comprised a 1-to-3 beam divider, three optical switches that were active Mach-Zehnder interferometers, and a 3-to-1 beam combiner, resulting in a 3-path interference. The three-arm interferometer could be used to test Born's rule—the correlation between the probability density and the wavefunction—in low Earth orbit (LEO). Another potential application for hBN quantum emitters is quantum memory. To be useful for quantum memory, the emitter-cavity system has to possess the following: a $\Lambda$ electronic structure, a compatible ZPL, a practical Q factor of a cavity, and a wide bandwidth. Cholsuk et al. identified 257 defect configurations with triplet-states, which possess $\Lambda$ energy levels, using DFT (**Figure 9b**, bottom panel).[109] Of these, 25 complexes closely matched the ZPLs of other quantum systems. In addition, most defect centers required a cavity Q-factor from $10^5$ to $10^7$ and received wide bandwidths to achieve 95% writing efficiency. The work showcased the potential of hBN quantum emitters for efficient quantum memory systems.

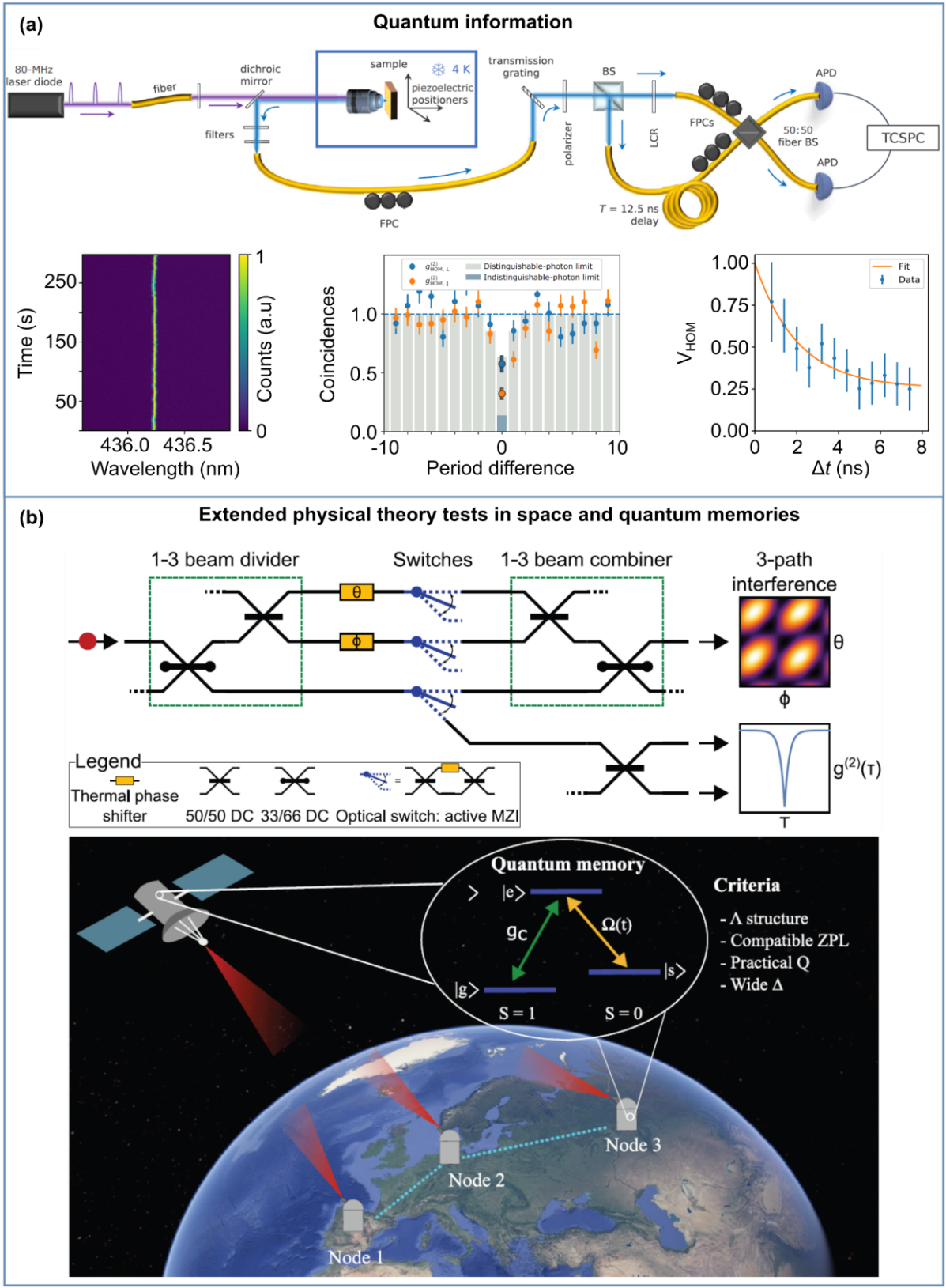

**Figure 9: Prospective approach for hBN quantum emitters. (a)** Quantum information with indistinguishability two-photon interference from quantum emitter generated from hBN via Hong-Ou-Mandel interferometer **(b)** Extended physical theory tests in space and quantum

memories with hBN quantum emitters. Figure a adapted with permission.[79] Copyright 2023, American Physical Society. Figure b adapted with permission.[108] Copyright 2024, Wiley-VCH GmbH.[109] Copyright 2024, Wiley-VCH GmbH.

## 11. Robustness of the quantum emitters

To enable the aforementioned applications, hBN quantum emitters must be robust towards environmental factors such as chemicals, ionized radiation, temperature, and mechanical stress. Quantum emitters were tested against gaseous substances such as argon, hydrogen, oxygen, and ammonia at 500°C for 1 hour (**Figure 10a**).[25] While some emitters bleached out, most survived the test with preserved photoluminescence spectra and quantum emission characteristics. Such a feat implied the excellent chemical resistance of the quantum emitters thanks to the high inertness of the hBN lattice structure. Next, the quantum emitters were subject to $\gamma$-ray irradiation to examine if they were space-qualified. Vogl and co-workers employed the isotope $^{22}_{11}Na$, which emitted 1.28 MeV photons and decayed to $^{22}_{10}Ne$, to study the robustness of the quantum emitters.[110] Interestingly, even with such highly energetic photons, hBN quantum emitters remained mostly intact, exhibiting the same PL lineshape and single-photon purity as shown in **Figure 10b**. To validate the high-temperature operation of quantum emitters in hBN, Kianinia et al. characterized the photophysics of the emitters in a home-built high-temperature PL chamber. They observed the emission characteristics of the quantum emitters from room temperature to 800 K, and negligible changes were observed for the photoluminescence, antibunching dip, and lifetimes (**Figure 10c**). The quantum emitters were also examined against tensile stress in an experimental attempt by Mendelson and co-workers.[111] The team observed that the emission was red-shifted significantly when the tensile strain was applied to the hBN film, which was attributed to perturbation onto the transition dipole moment. As shown in **Figure 10d**, reversible strain values as large as 5% were recorded during the experiment, indicating the high resistance of hBN towards strain-induced damage. Overall, quantum emitters have been shown to exhibit a high degree of robustness toward a variety of environmental elements and external fields, making them a promising candidate for ground-based and extraterrestrial quantum applications.

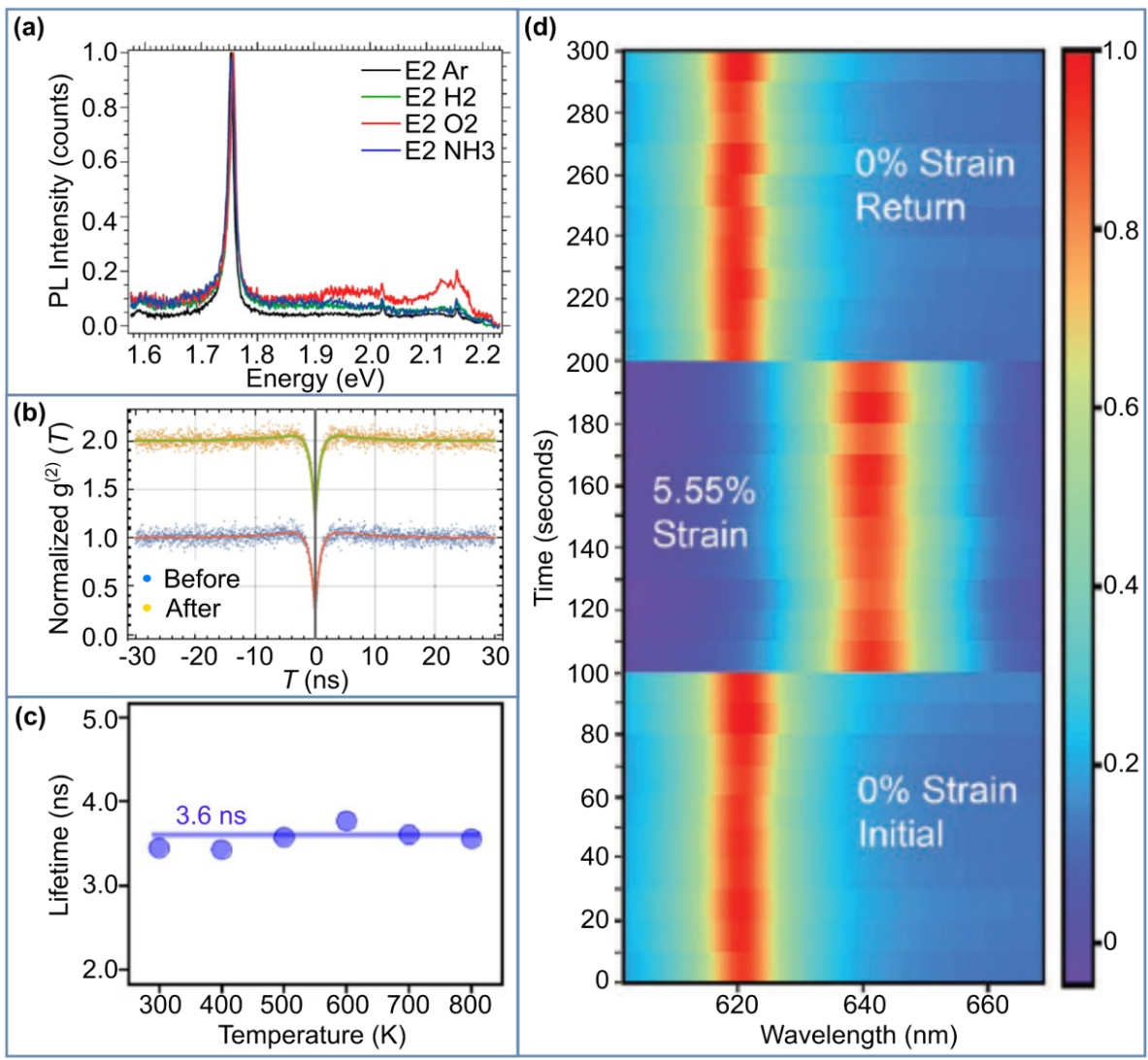

**Figure 10: Robustness of hBN quantum emitters in harsh conditions. (a)** Robust single photon emission at room temperature from hBN quantum emitters after the 30-minute sequential annealing in argon, hydrogen, oxygen and ammonia environments **(b)** Negligible differences in second-order correlation function dipping at zero time delay of hBN quantum emitter before (0.185) and after (0.188) gamma ray irradiation tolerance testing for space applications **(c)** Optically stable single photon source in hBN operating at high-temperature (800K) **(d)** Reversible tuning of the emission and related photophysical properties of hBN quantum emitters upon applying tensile strain. Figure a adapted with permission.[25] Copyright 2016, American Chemical Society. Figure b adapted under terms of the CC-BY 4.0 license.[110] Copyright 2019, The Authors, published by Springer Nature. Figure c adapted with permission.[112] Copyright 2017, American Chemical Society. Figure d adapted with permission.[111] Copyright 2020, Wiley-VCH Verlag GmbH & Co. KgaA, Weinheim.

## 12. Conclusion and Outlook

Eight years after its inception, quantum emitters in hBN have witnessed tremendous growth in interest from the research community worldwide. Such growing interest stems from the accessibility of the quantum material, which is inexpensive and readily available. Unlike other existing quantum sources, hBN quantum emitters feature a balanced suite of room-temperature operation, high brightness, good single-photon purity, high chemical robustness, spin-accessible (some families), and low fabrication costs. Being atomically thin and flat, hBN can be easily integrated into various photonic structures, monolithically or non-monolithically, from photonic crystal cavities and plasmonic antennas to hybrid architectures and integrated quantum photonic chips. Moreover, owing to their vicinity to the surface, hBN quantum emitters have garnered significant attention as frontrunners for the quantum sensing of various physical quantities and chemical species. Quantum key distribution, quantum communication, and physical theory testing in extraterrestrial contexts are other exciting applications for quantum emitters in hBN. Space qualification and other relevant testing and planning for the self-contained integrated quantum source have been established and are ready for the initial experiments in space. With the help of high-throughput DFT approaches, over two hundred defect complexes were proposed, many possessing the triplet-single intercrossing system critical for quantum sensing and quantum memory.

On the one hand, research into expanding the possible quantum applications for hBN emitters will be proliferating. Implementations such as quantum repeaters or teleportation are expected to be realized soon, considering the readiness of hBN emitters for such settings. On the other hand, the hunt for new defect families with superior quantum optical properties will continue with the help of efficient computational simulations. More unification between simulations and experiments, especially those with extreme resolution, such as TEM or STM, will help identify the chemical structures of these defect complexes. Furthermore, an electrically driven quantum source in hBN is another exciting research area—that is reachable thanks to the recent developments in two-dimensional heterostructure fabrication. Electrical excitation is essential for scalability since it can be implemented in complementary metal-oxide semiconductor technology. There are, however, several critical challenges remaining. First, although the ZPL distribution among emitters has been significantly improved with the discovery of the blue emitter family, such a parameter is still larger than competing platforms such as the NV or group IV centers in diamonds. Strain-engineering during or after growth may alleviate this heterogeneous linewidth broadening issue as it did in other platforms.[113] Second, spectral diffusion has been a long-standing problem for hBN emitters.[114] Such a phenomenon gives rise to low quantum interference visibility,[79] affecting many applications underpinned by the performance of this metric. Current directions include the preparation of ultrapure hBN surfaces/interfaces to minimize surface states, feedback-driven active tuning, or resonant excitation.[70] Third, due to its intrinsically large bandgap, making good electrical contact with hBN and introducing charges into the defect levels is incredibly challenging. Such a roadblock might be tackled by growing p-type and n-type hBN materials. While the former has been achieved, the latter has remained out of reach. Recent computational studies, however, shed light on the potential use of sacrificial impurity coupling to realize n-type hBN.[115] While it is a long journey ahead for the quantum source, it holds great promise to be a prime candidate for many quantum technologies of tomorrow.

## Acknowledgments

T. T. T acknowledges the financial support from the Australian Research Council (DE220100487, DP240103127). T. T. T and T. D. thank the Queensland Government for their financial support through the Quantum Challenge 2032 (Q2032).

## Conflict of Interest

The author declares no conflict of interest.

## ToC

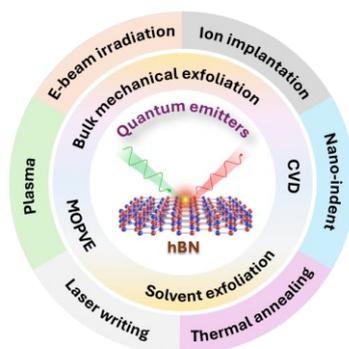

Quantum emitters in hexagonal boron nitride have emerged as a promising candidate for quantum information science. This review examines the fundamentals of these quantum emitters, including their level structures, defect engineering, and their possible chemical structures. It also explores the integration of these emitters into various photonic architectures and examines their current applications in quantum technologies.

First and corresponding author biographies and photographs

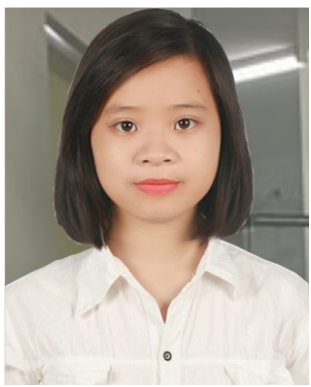

**Thi Ngoc Anh Mai** received her Bachelor degree in Physics and Education from the Hanoi National University of Education, Vietnam (2017) and Master of Science degree in Physics and Applications from the ENS Paris–Saclay, University of Paris–Saclay, France (2018). She worked as an university lecturer at the Faculty of Engineering Physics and Nanotechnology, VNU University of Engineering and Technology (VNU-UET), Hanoi, Vietnam (2019-2022). She is currently a Ph.D. student at the University of Technology Sydney, Australia (2022–Current). Her research interests focus on novel quantum emitters in wide-bandgap materials and their applications for quantum technologies.

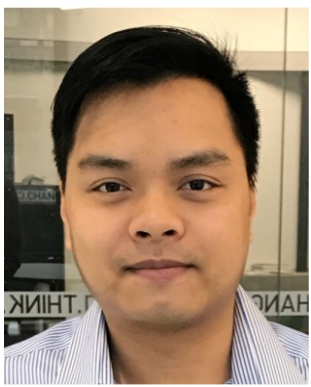

**Toan Trong Tran** is a Senior Lecturer (Associate Professor equivalent in the US) at the University of Technology Sydney (Australia). He received his Bachelor of Science in Materials Science from Vietnam National University, Ho Chi Minh City (VNU–HCMC), Vietnam (2008) and Master of Engineering in Chemical Engineering from National University of Singapore (NUS), Singapore (2011). He worked as a Research and Development Engineer at SDK Singapore (2011–2014). He received his PhD (2018) and worked as a Chancellor Postdoctoral Research Fellow (2018–2022) and a DECRA Fellow (2022—Current) and Senior Lecturer at the University of Technology Sydney, Australia. His research focuses on quantum optics, nanophotonics, solid–state physics, thermometry and nanofabrication.